\documentclass[10pt,conference]{IEEEtran}
\usepackage{amssymb}
\usepackage{amsmath}
\usepackage{cite}
\usepackage{url}
\usepackage{xcolor}
\usepackage{cite,graphicx,amsmath,amssymb}
\usepackage{subfigure}
\usepackage{cite}
\usepackage{caption}
\usepackage{amsthm}
\usepackage{algorithm}
\usepackage{algorithmic}
\allowdisplaybreaks

\usepackage[
top    = 0.75 in,
bottom = 1.05 in,
left   = 0.63 in,
right  = 0.63 in]{geometry}

\newtheorem{theorem}{Theorem}

\newtheorem{lemma}{Lemma}

\newtheorem{corollary}{Corollary}

\newcommand{\bm}[1]{\mbox{\boldmath{$#1$}}}

\usepackage{makecell}
\IEEEoverridecommandlockouts

\begin{document}
	
	\title{Realizing Fully-Connected Layers Over the Air via Reconfigurable Intelligent Surfaces}
	\author{
			\IEEEauthorblockN{Meng Hua, Chenghong~Bian, Haotian~Wu,  and   Deniz~G\"und\"uz}
		\IEEEauthorblockA{Department of Electrical and Electronic Engineering, Imperial College London, London SW7 2AZ, U.K}
		Email: {{\{m.hua, c.bian22, haotian.wu17, d.gunduz\}}@imperial.ac.uk}	
		\thanks{This work was  supported by the SNS JU Project 6G-GOALS under the	EU’s Horizon Program with Grant  101139232.
		}
	}

\maketitle
\begin{abstract}
By leveraging the waveform superposition property of the multiple access channel, over-the-air computation (AirComp) enables the execution of digital computations through analog means in the wireless domain, leading to faster processing and reduced latency. In this paper, we propose a novel approach to implement a neural network (NN) consisting of digital fully connected (FC) layers using physically reconfigurable hardware. Specifically, we investigate reconfigurable intelligent surfaces (RISs)-assisted multiple-input multiple-output (MIMO) systems to emulate the functionality of a NN for over-the-air inference. In this setup, both the RIS and the transceiver are jointly configured to manipulate the ambient wireless propagation environment, effectively reproducing the adjustable weights of a digital FC layer. We refer to this new computational paradigm as \textit{AirFC}. We formulate an imitation error minimization problem between the effective channel created by RIS  and a target  FC layer by jointly optimizing over-the-air parameters. To solve this non-convex optimization problem,  an extremely low-complexity alternating optimization algorithm is proposed, where semi-closed-form/closed-form solutions for all optimization variables are derived. Simulation results show that the RIS-assisted MIMO-based AirFC can achieve competitive classification accuracy. Furthermore, it is also shown that a multi-RIS configuration significantly outperforms a single-RIS setup, particularly in line-of-sight (LoS)-dominated channels.
\end{abstract}

\section{Introduction}
Machine learning (ML)  is typically trained in the digital domain using graphics-processing units, which are time- and energy-intensive due to the separation of memory and processing in von Neumann architectures. To address this limitation, one promising approach is to implement neural networks (NNs) using physical hardware, such as optics, mechanics, and electronics, operating in the analog domain. These are termed physical neural networks (PNNs) \cite{wright2022deep,hughes2019wave,van2023retrainable}. This approach enables the training of deep PNNs composed of controllable physical layers, where computations typically performed in the digital domain are approximated by the propagation of waves or optical signals through physical media. Since such computations occur over the air at the speed of light, this method offers significantly reduced latency and power consumption. Inspired by this principle, the use of multiple-input-multiple-output (MIMO) over-the-air computation (AirComp)  for realizing fully connected (FC)  layers with fast and green computation is possible. This advantage arises because AirComp leverages the superposition property of the wireless channels to inherently perform  algebraic functions \cite{sahin2023survey,zhao2023Intelligent,zhu20196mimo}.
 
Currently, only a few studies have explored using AirComp to implement FC layers  \cite{reus2023airfc,yang2023over}.  In \cite{reus2023airfc}, a linear function in the FC layer is realized through the superposition of orthogonal frequency-division multiplexing (OFDM) signals emitted by multiple transmitters and received by a single receiver, i.e., a multiple-input single-output (MISO) system.  This approach was extended to a MIMO system in \cite{yang2023over}, where both the precoder and combiner are trained jointly over the air.  However, these methods design the transmission waveforms to adapt to the specific channel, implying that system performance is highly sensitive to channel conditions such as fading and rank.

To address these challenges, this paper proposes using multiple reconfigurable intelligent surfaces (RISs) to enhance channel conditions  \cite{pan2021Reconfigurable,Hua2024Intelligent,chen2022active,meng2022Intelligent}.  Specifically, we investigate a multi-RIS-aided MIMO AirComp system that emulates a digital FC layer. In this system, both RISs and the transceiver are jointly adjusted to shape the wireless environment, effectively realizing digital layer weights in the analog domain. We refer to this new computational model as \textit{AirFC}. 
We formulate the problem as minimizing the imitation error between the AirFC system and a given digital FC layer by jointly optimizing all over-the-air parameters. To solve this non-convex optimization problem, we propose an efficient alternating optimization algorithm that partitions all variables into three blocks and optimizes them alternately until convergence. In particular, a closed-form solution for the precoder is derived via the Lagrange duality method, while semi-closed-form solutions for the combiner and RIS phase shifts are also obtained.
Simulation results demonstrate that the RIS-aided MIMO AirComp system can effectively replicate the behavior of a digital FC layer. Additionally, significant classification accuracy improvements can be achieved  by the multi-RIS setup.

\begin{figure}[!t]
	\centerline{\includegraphics[width=3.1in]{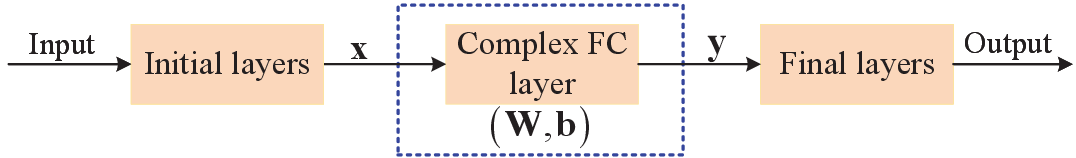}}
	\caption{Digital-domain-based NN architecture.} \label{model_fig1}
\end{figure}
\begin{figure}[!t]
	\centerline{\includegraphics[width=3.5in]{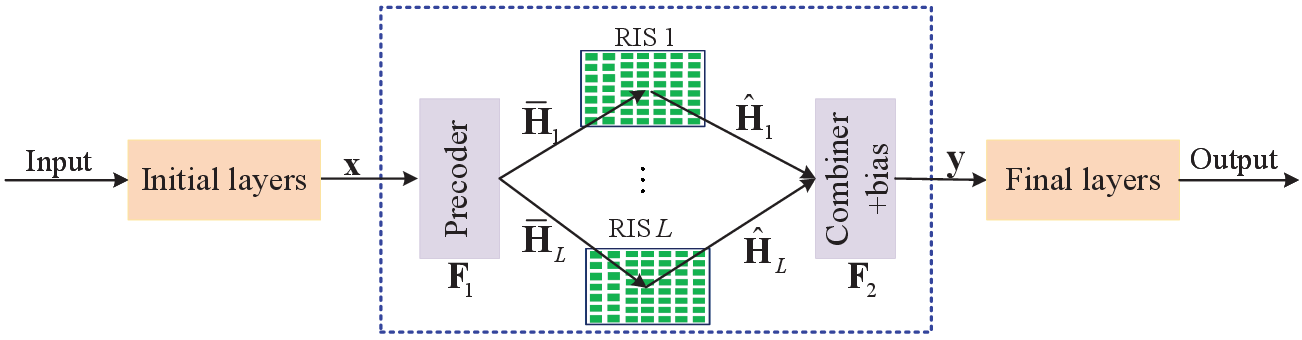}}
	\caption{Analog-domain-based NN  architecture.} \label{model_fig1_1}
\end{figure}
\section{System Model and Problem Formulation}
\subsection{System Model}
Fig.~\ref{model_fig1} depicts a conventional digital-domain-based complex-valued NN architecture divided into three components:  initial layers, a middle layer, and final layers. The middle layer is a digital complex FC layer, while the initial and final layers may contain numerous functional components such as residual blocks, transformer layers, multi-head attention mechanisms, etc., depending on specific applications. Let ${\bf{x}} \in {{\mathbb C}^{N \times 1}}$  and ${\bf{y}} \in {{\mathbb C}^{N \times 1}}$ represent the input and output of the complex FC layer, respectively, where $N$ is the dimensionality of both the input and output. The FC layer computation can then be expressed as
\begin{align}
	{\bf{y}} = {\bf{Wx}} + {\bf{b}}, \label{linearfunction}
\end{align}
where  ${\bf{W}} \in {{\mathbb C}^{N \times N}}$ and ${\bf{b}} \in {{\mathbb C}^{N \times 1}}$ denote the weight matrix and bias vector of the FC layer, respectively.

Fig.~\ref{model_fig1_1} illustrates an over-the-air analog-domain-based NN architecture that also consists of three components: initial layers, a transmission layer, and final layers. 
The transmission layer comprises three components, namely a precoder, a combiner, and multiple RISs.   We assume that the number of transmit and receive antennas is  $N$, matching the input/output dimensions of the digital FC layer described in Fig.~\ref{model_fig1}. In addition, we assume that there are $L$  RISs deployed  separately, with the $i$th RIS containing  $M_i$ 
reflecting elements,   satisfying $\sum\limits_{i = 1}^L {{M_i} = M} $, and   $M$ denotes the total  number of reflecting elements across $L$ RISs. Without loss of generality, we set $M_i=M/L$ for $i = 1, \ldots ,L$.
Let ${{\bf{\bar H}}_i} \in {{\mathbb C}^{\frac{M}{L} \times N}}$ and ${{\bf{\hat H}}_i} \in {{\mathbb C}^{N \times \frac{M}{L}}}$ denote the complex equivalent baseband channel matrices from the transmitter to  RIS $i$, and from  RIS $i$ to the receiver, respectively. In addition, the  phase-shift matrix  for RIS $i$ is denoted by ${{\bf{\Theta }}_i} = {\rm{diag}}\left( {{e^{j{\theta _{i,1}}}},{e^{j{\theta _{i,2}}}}, \ldots ,{e^{j{\theta _{i,M/L}}}}} \right)$, $i \in \left\{ {1, \ldots ,L} \right\}$, where $\theta_{i,m}$ denotes the $m$th  phase shift in RIS $i$. For simplicity and clarity, we assume that the direct link between the transmitter and the receiver is blocked. Nevertheless, the proposed algorithm is also applicable in scenarios where a direct path is present.

\subsection{Channel Rank Improvement via Multi-RIS}
To better illustrate the impact of channel rank on imitation performance, we first consider a scenario involving a single RIS with $M$ reflecting elements. Denote by ${{{\bf{ H}}}_1}\in {{\mathbb C}^{M \times N}}$ and ${{{\bf{ H}}}_2}\in {{\mathbb C}^{N \times M}}$ the complex equivalent baseband channels from the transmitter to the RIS and from the RIS to  the receiver, respectively.
If $\bf W$ is a high-rank matrix, while channels ${\bf H}_1$
and ${\bf H}_2$ are line-of-sight (LoS), we have
\begin{align}
{\rm{rank}}\left( {{{\bf{H}}_2}{\bf{\Theta }}{{\bf{H}}_1}} \right) &= \min \left\{ {{\rm{rank}}\left( {{{\bf{H}}_1}} \right),{\rm{rank}}\left( {{{\bf{H}}_2}} \right)} \right\}=1, \notag\\
&\ll {\rm{rank}}\left( {\bf{W}} \right), \label{oneRIS: rank}
\end{align}
then the product  ${{{\bf{F}}_2}{{\bf{H}}_2}{\bf{\Theta }}{{\bf{H}}_1}{{\bf{F}}_1}}$ ( with ${{{\bf{F}}_1}}$ and ${{{\bf{F}}_2}}$ denoting the transmit precoder and the receive combiner, respectively) cannot match the rank of ${\bf{W}}$, regardless of how the over-the-air parameters  ${\bf F}_1$, ${\bf F}_2$, and $\bf \Theta$ are optimized. 
To solve this rank deficiency issue, we propose deploying multiple RISs.
The effective over-the-air channel for a multi-RIS setup is given by (ignoring higher-order reflections due to significant path loss)
\begin{align}
{\bf{H}} = \sum\limits_{i = 1}^L {{{{\bf{\hat H}}}_i}{{\bf{\Theta }}_i}{{{\bf{\bar H}}}_i}} 
= {\bf{\hat H}}{\rm{diag}}\left( {{{\bf{\Theta }}_1}, \ldots ,{{\bf{\Theta }}_L}} \right){\bf{\bar H}}, \label{channel:multiRIS}
\end{align}
where ${\bf{\hat H}} = \left[ {{{{\bf{\hat H}}}_1}, \ldots ,{{{\bf{\hat H}}}_L}} \right]$ and ${\bf{\bar H}} = {\left[ {{\bf{\bar H}}_1^T, \ldots ,{\bf{\bar H}}_L^T} \right]^T}$.

Although  \eqref{channel:multiRIS} resembles the single-RIS case in form,  its rank properties differ. Specifically, we have the inequality:
\begin{align}
{\rm{rank}}\left( {\bf{H}} \right) \le \min \left\{ {{\rm{rank}}\left( {\sum\limits_{i = 1}^L {{{{\bf{\hat H}}}_i}} } \right),{\rm{rank}}\left( {\sum\limits_{i = 1}^L {{{{\bf{\bar H}}}_i}} } \right)} \right\}. \label{multiRIS: inequaity1}
\end{align}
In the worst-case, where all channels are LoS, i.e, ${\rm{rank}}\left( {{{{\bf{\hat H}}}_i}} \right) = {\rm{rank}}\left( {{{{\bf{\bar H}}}_i}} \right) = 1$,   $i \in \left\{ {1, \ldots ,L} \right\}$, and $L$ RISs are physically separated, the ranks of the aggregated channel matrices satisfy
\begin{align}
{\rm{rank}}\left( {\sum\limits_{i = 1}^L {{{{\bf{\hat H}}}_i}} } \right) = L,~~ {{\rm{rank}}\left( {\sum\limits_{i = 1}^L {{{{\bf{\bar H}}}_i}} } \right)}=L. \label{rank:aggregated channel}
\end{align}
Substituting \eqref{rank:aggregated channel} into \eqref{multiRIS: inequaity1}, we obtain 
\begin{align}
{\rm{rank}}\left( {\bf{H}} \right) \le L,
\end{align}
indicating that with a sufficiently large  $L$, the rank of the channel $\bf H$ approximate that of $\bf W$. Therefore, by jointly optimizing ${\bf F}_1$, ${\bf F}_2$, and $\bf \Theta$, the imitation error is expected to be reduced significantly.

\subsection{Problem Formulation}
As shown in Fig.~\ref{model_fig1_1}, the  signal received at the receiver, including the bias,  is given by
\begin{align}
{\bf{y}} &={{\bf{F}}_2}\left( {\sum\limits_{i = 1}^L {{{{\bf{\hat H}}}_i}{{\bf{\Theta }}_i}{{{\bf{\bar H}}}_i}} {{\bf{F}}_1}{\bf{x}} + {\bf{n}}} \right) + {\bf{b}} \notag\\
&={{\bf{F}}_2}\left( {{\bf{\hat H\Theta }}{{{\bf{\bar H}}}}{{\bf{F}}_1}{\bf{x}} + {\bf{n}}} \right) + {\bf{b}},\label{channel:singleRIS}
\end{align} 
 where $\bf x$ represents input to the FC layer,  ${{{\bf{F}}_1}}$ and ${{{\bf{F}}_2}}$ denote the transmit precoder and  receive combiner, respectively, and  ${\bf{\Theta }} = {\rm{diag}}\left( {{{\bf{\Theta }}_1}, \ldots ,{{\bf{\Theta }}_L}} \right)$.  The term ${\bf{n}} \sim {\cal CN}\left( {0,{\sigma ^2{\bf I}_N}} \right)$ is the additive
white Gaussian noise, and ${\bf{b}}$ is the bias term defined in \eqref{linearfunction}.

In this paper, we aim to use the transmission layer in Fig.~\ref{model_fig1_1} to imitate the behavior of the complex FC layer in Fig.~\ref{model_fig1} via jointly designing the precoder, the combiner, and the multi-RIS phase shifts.  The  imitation minimization error optimization problem is formulated as follows:
\begin{subequations}  \label{P1}
\begin{align}
    &\mathop {\min }\limits_{{{\bf{F}}_1},{{\bf{F}}_2},{\bf{\Theta }}} \left\| {{{\bf{F}}_2}{{\bf{\hat H}}}{\bf{\Theta }}{{\bf{\bar H}}}{{\bf{F}}_1} - {\bf{W}}} \right\|_F^2 + {{\mathbb E}_{\bf{n}}}\left\{ {{{\left\| {{{\bf{F}}_2}{\bf{n}}} \right\|}^2}} \right\} \label{objectivefunction}\\
   & {\rm{s}}{\rm{.t}}{\rm{. }}~\left\| {{{\bf{F}}_1}} \right\|_F^2 \le {P_{{\rm{max}}}},\label{transmitpower}\\
   &\qquad \left| {{{\bf{\Theta }}_{i,i}}} \right| = 1,{\kern 1pt} {\kern 1pt} {\kern 1pt} {\kern 1pt} {\kern 1pt} {\kern 1pt} {\kern 1pt} i = 1, \ldots ,M.\label{RIS_phaseshift}
\end{align}
\end{subequations}
The first term in \eqref{objectivefunction} aims to minimize the imitation error in approximating the FC layer weights, while the second term addresses noise propagation through the bias. Constraint \eqref{transmitpower} limits the transmit power to not exceed $P_{\rm max}$, and \eqref{RIS_phaseshift}  enforces the unit-modulus constraint on each RIS phase shift element. 

Problem \eqref{P1} is inherently non-convex due to the coupling of the optimization variables in the objective function \eqref{objectivefunction}, and the non-convex unit-modulus constraint \eqref{RIS_phaseshift}. 
There are no standard methods for optimally solving such problems in general. To address this, a low-complexity alternating optimization algorithm is proposed in Section \ref{section:non-trainable}.

\section{Proposed Solution}\label{section:non-trainable}
In this section, we propose an efficient algorithm to solve problem \eqref{P1} using an alternating optimization method. Specifically, we partition all optimization variables into three blocks, namely ${{{\bf{F}}_1}}$, ${{{\bf{F}}_2}}$, and ${\bf{\Theta }}$, and then alternately optimize each block until convergence is achieved.
\subsubsection{For any given RIS phase shift ${\bf{\Theta }}$ and receive combiner ${{{\bf{F}}_2}}$, the subproblem corresponding to the transmit precoder optimization is given by (ignoring  irrelevant constants)}
\begin{subequations}  \label{P1_transmitbeamformer}
	\begin{align}
	&\mathop {\min }\limits_{{{\bf{F}}_1}} \left\| {{{\bf{F}}_2}{{\bf{\hat H}}}{\bf{\Theta }}{{\bf{\bar H}}}{{\bf{F}}_1} - {\bf{W}}} \right\|_F^2 \label{transmitbeamformer_objectivefunction}\\
	& {\rm{s}}{\rm{.t}}{\rm{. }}~\eqref{transmitpower}.
	\end{align}
\end{subequations}
It can be readily checked that problem \eqref{P1_transmitbeamformer} is a quadratically
constrained quadratic program (QCQP). In the following, 
we derive a semi-closed-form yet optimal solution for problem \eqref{P1_transmitbeamformer}  by using the Lagrange duality method \cite{boyd2004convex}. To  be specific, by introducing dual variable $\lambda \ge 0 $ with constraint \eqref{transmitpower},  the Lagrangian function of
problem  \eqref{P1_transmitbeamformer} is given by 
\begin{align}
{\cal L}\left( {{{\bf{F}}_1},\lambda } \right) = \left\| {{{\bf{F}}_2}{{\bf{\hat H}}}{\bf{\Theta }}{{\bf{\bar H}}}{{\bf{F}}_1} - {\bf{W}}} \right\|_F^2 + \lambda \left( {\left\| {{{\bf{F}}_1}} \right\|_F^2 - {P_{{\rm{max}}}}} \right).
\end{align}
By taking the  first-order derivative of ${\cal L}\left( {{{\bf{F}}_1},\lambda } \right)$ with ${\bf{F}}_1^*$ and setting it to zero, we obtain the optimal solution of ${{{\bf{F}}_1}}$ as 
 \begin{align}
 {\bf{F}}_1^{{\rm{opt}}} = {\left( {{{\bm \Upsilon} ^H}{\bm \Upsilon}   + \lambda {{\bf{I}}_N}} \right)^{ - 1}}{{\bm \Upsilon}  ^H}{\bf{W}}, \label{beamformer}
 \end{align}
where ${\bm \Upsilon}   = {{\bf{F}}_2}{{\bf{\hat H}}}{\bf{\Theta }}{{\bf{\bar H}}}$. According to the complementary slackness condition \cite{boyd2004convex}, the optimal solutions $ {\bf{F}}_1^{{\rm{opt}}}$ and ${\lambda ^{{\rm{opt}}}}$ should satisfy the following equation:
 \begin{align}
 {\lambda ^{{\rm{opt}}}}\left( {\left\| {{\bf{F}}_1^{{\rm{opt}}}} \right\|_F^2 - {P_{{\rm{max}}}}} \right) = 0.
 \end{align}
There are two possible solutions: 1) ${\lambda ^{{\rm{opt}}}}=0$ and ${\left\| {{\bf{F}}_1^{{\rm{opt}}}} \right\|_F^2 \le {P_{{\rm{max}}}}}$; 2) ${\lambda ^{{\rm{opt}}}}>0$ and ${\left\| {{\bf{F}}_1^{{\rm{opt}}}} \right\|_F^2 ={P_{{\rm{max}}}}}$. We first check whether ${\lambda ^{{\rm{opt}}}}=0$ is the optimal solution or not. If 
\begin{align}
\!\left\| {{\bf{F}}_1} \right\|_F^2 - {P_{{\rm{max}}}} = \left\| {{{\left( {{{\bm \Upsilon}  ^H}{\bm \Upsilon} } \right)}^{ - 1}}{{\bm \Upsilon} ^H}{\bf{W}}} \right\|_F^2 - {P_{{\rm{max}}}} < 0,
\end{align}
which indicates that  ${\bf{F}}_1^{{\rm{opt}}} = {\left( {{{\bm \Upsilon}^H}{\bm \Upsilon} } \right)^{ - 1}}{{\bm \Upsilon} ^H}{\bf{W}}$; otherwise, we should calculate $\lambda $ that renders $\left\| {{{\left( {{{\bm \Upsilon} ^H}{\bm \Upsilon}  + \lambda {{\bf{I}}_N}} \right)}^{ - 1}}{{\bm \Upsilon} ^H}{\bf{W}}} \right\|_F^2 = {P_{{\rm{max}}}}$. It can be readily checked that ${{\bm \Upsilon} ^H}{\bm \Upsilon}$ is  a positive semi-definite matrix, we can perform eigendecomposition on ${{\bm \Upsilon} ^H}{\bm \Upsilon}$, and yield 
\begin{align}
{{\bm \Upsilon} ^H}{\bm \Upsilon}  = {\bf{U\Sigma }}{{\bf{U}}^H}, \label{eigendecomposition}
\end{align}
where ${\bf{U}}{{\bf{U}}^H} = {{\bf{U}}^H}{\bf{U}} = {{\bf{I}}_N}$ and ${\bf{\Sigma }}$ is a diagonal matrix consisting of eigenvalues.
 Substituting \eqref{eigendecomposition} into $\left\| {{\bf{F}}_1^{{\rm{opt}}}} \right\|_F^2$ in \eqref{beamformer},  we have 
 \begin{align}
 \left\| {{\bf{F}}_1^{{\rm{opt}}}} \right\|_F^2 &= {\rm{tr}}\left( {{{\left( {{\bf{\Sigma }} + \lambda {{\bf{I}}_N}} \right)}^{ - 2}}{{\bf{U}}^H}{{\bm \Upsilon} ^H}{\bf{W}}{{\bf{W}}^H}{\bm \Upsilon} {\bf{U}}} \right)\notag\\
& = \sum\limits_{i = 1}^N {\frac{{{{\left( {{{\bf{U}}^H}{{\bm \Upsilon}  ^H}{\bf{W}}{{\bf{W}}^H}{\bm \Upsilon}  {\bf{U}}} \right)}_{i,i}}}}{{{{\left( {{{\bf{\Sigma }}_{i,i}} + \lambda } \right)}^2}}}}.
 \end{align}
 It is not difficult to verify that $\left\| {{\bf{F}}_1^{{\rm{opt}}}} \right\|_F^2$ is monotonically decreasing with $\lambda $, and a simple bisection method can be applied for solving it. To confine the search range of $\lambda $, the upper bound of $\lambda $ can be set as 
 \begin{align}
 {\lambda ^{{\rm{up}}}} = \sqrt {\frac{{\sum\limits_{i = 1}^N {{{\left( {{{\bf{U}}^H}{{\bm \Upsilon}  ^H}{\bf{W}}{{\bf{W}}^H}{\bm \Upsilon}  {\bf{U}}} \right)}_{i,i}}} }}{{{P_{{\rm{max}}}}}}}. 
 \end{align}
 Thus, the search range of $\lambda $ is given by $\left[ {0,{\lambda ^{{\rm{up}}}}} \right]$.

\subsubsection{For any given RIS phase shift ${\bf{\Theta }}$ and transmit precoder  ${{{\bf{F}}_1}}$, the subproblem corresponding to the receive combiner optimization is given by (ignoring  irrelevant constants)}

\begin{subequations}  \label{P1_receivecombiner}
	\begin{align}
	&\mathop {\min }\limits_{{{\bf{F}}_2}} \left\| {{{\bf{F}}_2}{{\bf{\hat H}}}{\bf{\Theta }}{{\bf{\bar H}}}{{\bf{F}}_1} - {\bf{W}}} \right\|_F^2 + {{\mathbb E}_{\bf{n}}}\left\{ {{{\left\| {{{\bf{F}}_2}{\bf{n}}} \right\|}^2}} \right\}. \label{receivecombiner_objectivefunction}
	\end{align}
\end{subequations}
Note that ${{\mathbb E}_{\bf{n}}}\left\{ {{{\left\| {{{\bf{F}}_2}{\bf{n}}} \right\|}^2}} \right\} = {\sigma ^2}{\rm{tr}}\left( {{{\bf{F}}_2}{\bf{F}}_2^H} \right)$, objective function  \eqref{receivecombiner_objectivefunction} is thus quadratic and its global solution can be attained. Specifically,  taking the  first-order derivative of $\left\| {{{\bf{F}}_2}{{\bf{\hat H}}}{\bf{\Theta }}{{\bf{\bar H}}}{{\bf{F}}_1} - {\bf{W}}} \right\|_F^2 + {\sigma ^2}{\rm{tr}}\left( {{{\bf{F}}_2}{\bf{F}}_2^H} \right)$ with ${\bf{F}}_1^*$ and setting it to zero, its optimal solution of ${{{\bf{F}}_2}}$ is given by 
\begin{align}
{{\bf{F}}_2^{\rm opt}} = {\left( {\bar {\bm \Upsilon} {{\bar {\bm \Upsilon} }^H} + {\sigma ^2}{{\bf{I}}_N}} \right)^{ - 1}}{\bf{W}}{{\bar {\bm \Upsilon} }^H}, \label{combiner:optimal}
\end{align}
where $\bar {\bm \Upsilon} = {{\bf{\hat H}}}{\bf{\Theta }}{{\bf{\bar H}}}{\bf{F}}_1$.

\subsubsection{For any given transmit precoder  ${{{\bf{F}}_1}}$ and receive combiner ${{{\bf{F}}_2}}$, the subproblem corresponding to the RIS phase shift ${\bf{\Theta }}$  optimization is given by (ignoring  irrelevant constants)}
\begin{subequations}  \label{P1_RIS}
	\begin{align}
	&\mathop {\min }\limits_{{\bf{\Theta }}} \left\| {{{\bf{F}}_2}{{\bf{\hat H}}}{\bf{\Theta }}{{\bf{\bar H}}}{{\bf{F}}_1} - {\bf{W}}} \right\|_F^2 \label{RIS_objectivefunction}\\
	& {\rm{s}}{\rm{.t}}{\rm{. }}~\eqref{RIS_phaseshift}.
	\end{align}
\end{subequations}
Problem \eqref{P1_RIS} is non-convex due to the non-convexity of constraint \eqref{RIS_phaseshift}. To solve it, we first unfold the objective function \eqref{RIS_objectivefunction} as 
\begin{align}
    &\left\| {{{\bf{F}}_2}{{\bf{\hat H}}}{\bf{\Theta }}{{\bf{\bar H}}}{{\bf{F}}_1} - {\bf{W}}} \right\|_F^2 = \notag\\
    &{\rm{tr}}\left( {{{\bf{F}}_2}{{\bf{\hat H}}}{\bf{\Theta }}{{\bf{\bar H}}}{{\bf{F}}_1}{\bf{F}}_1^H{\bf{H}}_1^H{{\bf{\Theta }}^H}{\bf{F}}_2^H{\bf{H}}_2^H} \right) + {\rm{tr}}\left( {{\bf{W}}{{\bf{W}}^H}} \right)-\notag\\
 &{\rm{tr}}\left( {{{\bf{F}}_2}{{\bf{\hat H}}}{\bf{\Theta }}{{\bf{\bar H}}}{{\bf{F}}_1}{{\bf{W}}^H}} \right) - {\rm{tr}}\left( {{\bf{WF}}_1^H{\bf{\bar H}}^H{{\bf{\Theta }}^H}{\bf{F}}_2^H{\bf{\hat H}}^H} \right). \label{equation_1}
\end{align}
Then, we can rewrite  terms in \eqref{equation_1} into a more compact form given by
\begin{align}
    &{\rm{tr}}\left( {{{\bf{F}}_2}{{\bf{\hat H}}}{\bf{\Theta }}{{\bf{\bar H}}}{{\bf{F}}_1}{\bf{F}}_1^H{\bf{\bar H}}^H{{\bf{\Theta }}^H}{\bf{F}}_2^H{\bf{\hat H}}^H} \right)  \notag\\
    &= {\rm{tr}}\left( {{{\bf{\Theta }}^H}{\bf{F}}_2^H{\bf{\hat H}}^H{{\bf{F}}_2}{{\bf{\hat H}}}{\bf{\Theta }}{{\bf{\bar H}}}{{\bf{F}}_1}{\bf{F}}_1^H{\bf{\bar H}}^H} \right)\notag\\
    &  = {{\bf{v}}^H}{\bm \Omega} {\bf{v}}, \label{equation_2}
\end{align}
and
\begin{align}
{\rm{tr}}\left( {{{\bf{F}}_2}{{\bf{\hat H}}}{\bf{\Theta }}{{\bf{\bar H}}}{{\bf{F}}_1}{{\bf{W}}^H}} \right)& = {\rm{tr}}\left( {{\bf{\Theta }}{{\bf{\bar H}}}{{\bf{F}}_1}{{\bf{W}}^H}{{\bf{F}}_2}{{\bf{\hat H}}}} \right)\notag\\
& = {{\bf{v}}^T}{\bm{\varphi }}, \label{equation_3}
\end{align}
where  ${\bf{v}} = {\left( {{{\bf{\Theta }}_{1,1}}, \ldots ,{{\bf{\Theta }}_{M,M}}} \right)^T}$, ${\bm \Omega } = \left( {{\bf{F}}_2^H{\bf{\hat H}}^H{{\bf{F}}_2}{{\bf{\hat H}}}} \right) \odot {\left( {{{\bf{\bar H}}}{{\bf{F}}_1}{\bf{F}}_1^H{\bf{\bar H}}^H} \right)^T}$, and 
${\bm{\varphi }} = {\left[ {{{\left[ {{{\bf{\bar H}}}{{\bf{F}}_1}{{\bf{W}}^H}{{\bf{F}}_2}{{\bf{\hat H}}}} \right]}_{1,1}}, \ldots ,{{\left[ {{{\bf{\bar H}}}{{\bf{F}}_1}{{\bf{W}}^H}{{\bf{F}}_2}{{\bf{\hat H}}}} \right]}_{M,M}}} \right]^T}$.
Based on  \eqref{equation_2} and  \eqref{equation_3}, problem \eqref{P1_RIS} 
can be rewritten as 
\begin{subequations}  \label{P1_RIS_new}
	\begin{align}
	&\mathop {\min }\limits_{\bf{v}} {{\bf{v}}^H}{\bm \Omega} {\bf{v}} - 2{\mathop{\rm Re}\nolimits} \left\{ {{{\bf{v}}^T}{\bm{\varphi }}} \right\} + {\rm{tr}}\left( {{\bf{W}}{{\bf{W}}^H}} \right) \label{RIS_objective_newfunction}\\
	& {\rm{s}}{\rm{.t}}{\rm{. }}~\left| {{{\bf{v}}_i}} \right| = 1,i = 1, \ldots ,M.
	\end{align}
\end{subequations}
To solve problem \eqref{P1_RIS_new}, a majorization-minimization method can be applied \cite{sone2016sequence}. The key   idea behind the majorization-minimization method is to solve   problem \eqref{P1_RIS_new} by constructing a series of more tractable approximate
objective functions. Specifically,  ${{\bf{v}}^H}{\bm \Omega} {\bf{v}}$ is upper bounded by 
\begin{align}
{{\bf{v}}^H}{\bm \Omega} {\bf{v}} &\le {\lambda _{\max }}M - 2{\mathop{\rm Re}\nolimits} \left\{ {{{\bf{v}}^H}\left( {{\lambda _{\max }}{{\bf{I}}_M} - {\bm \Omega}  } \right){{\bf{v}}^r}} \right\}\notag\\
&+ {{\bf{v}}^{r,H}}\left( {{\lambda _{\max }}{{\bf{I}}_M} - {\bm \Omega}  } \right){{\bf{v}}^r}, \label{phaseshift_upperbound}
\end{align}
where ${{{\bf{v}}^r}}$ represents any initial point of ${{{\bf{v}}}}$. Substituting \eqref{phaseshift_upperbound} into \eqref{RIS_objective_newfunction}, problem \eqref{P1_RIS_new} is approximated as (ignoring irrelevant constants)
\begin{subequations}  \label{P1_RIS_new_approx}
	\begin{align}
	&\mathop {\min }\limits_{\bf{v}}  - 2{\rm{Re}}\left\{ {{{\bf{v}}^H}\left( {\left( {{\lambda _{\max }}{{\bf{I}}_M} - {\bm \Omega} } \right){{\bf{v}}^r} - {{\bm \varphi} ^*}} \right)} \right\} \\
	& {\rm{s}}{\rm{.t}}{\rm{. }}~\eqref{phaseshift_upperbound}.
	\end{align}
\end{subequations}
It is not difficult to prove  that the optimal solution to problem \eqref{P1_RIS_new_approx} is given by 
\begin{align}
	{{\bf{v}}^{{\rm{opt}}}} = {e^{j\arg \left( {\left( {{\lambda _{\max }}{{\bf{I}}_M} - {\bm \Omega} } \right){{\bf{v}}^r} - {{\bm \varphi} ^*}} \right)}}. \label{RISphaseshit_new}
\end{align}

\subsubsection{Overall algorithm}
\begin{algorithm}[!t]
	\caption{Alternating optimization  for solving  problem \eqref{P1}.}	\label{alg1}
	\begin{algorithmic}[1]
		\STATE  \textbf{Initialize} RIS phase-shift vector  ${{\bf{v}}}$ and combiner ${\bf F}_2$.
		\STATE  \textbf{repeat}
		\STATE  \quad Update transmit precoder ${\bf F}_1$ by solving  \eqref{P1_transmitbeamformer}.
		\STATE  \quad Update receive combiner ${\bf F}_2$  based on \eqref{combiner:optimal}
		\STATE \quad Update RIS phase shift  ${{\bf{v}}}$ based on \eqref{RISphaseshit_new}.
		\STATE \textbf{until}   the  decrease  in the objective function of problem \eqref{P1} falls below a predefined threshold.
	\end{algorithmic}
\end{algorithm}

Based on the solutions to the above subproblems, we solve each subproblem iteratively, where the solution obtained in the current iteration serves as the initial point for the next iteration, until convergence is reached. The detailed procedure is summarized in Algorithm~\ref{alg1}. 

The computational complexity of Algorithm~\ref{alg1} is analyzed  as follows: In step $3$, the complexity of calculating the precoder  ${\bf F}_1$ via the Lagrange duality method is ${\cal O}\left( {{{\log }_2}\left( {\frac{{{\lambda ^{{\rm{up}}}}}}{\varepsilon }{N^3}{M^3}} \right)} \right)$, where $\varepsilon $ is the predefined accuracy. In  step $4$, a closed-form expression for ${\bf F}_2$ is derived and its computational complexity is given by ${\cal O}\left( {{N^3}{M^3}} \right)$.  In  step $5$, a closed-form expression for ${\bf v}$ is derived and its computational complexity is given by ${\cal O}\left( {{M^3}} \right)$. Therefore, 
 the total complexity of Algorithm~\ref{alg1} is given by ${\cal O}\left( {{L_{{\rm{iter}}}}\left( {{{\log }_2}\left( {\frac{{{\lambda ^{{\rm{up}}}}}}{\varepsilon }} \right){N^3}{M^3} + {N^3}{M^3} + {M^3}} \right)} \right)$, where ${{L_{{\rm{iter}}}}}$ denotes the number of iterations required for convergence.


\begin{figure*}[!t]
	\centerline{\includegraphics[width=7.2in]{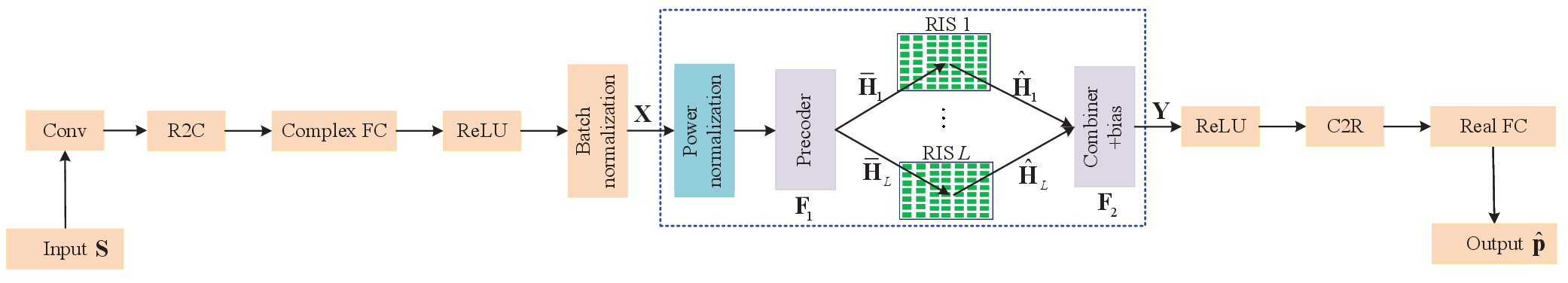}}
	\caption{A multi-RIS aided over-the-air transmission architecture.} \label{model_fig:multiRIS}
\end{figure*}

\section{Numerical Results}
In this section, we present numerical results to evaluate the performance of the proposed AirFC framework via  RIS.   To demonstrate the potential of RIS-aided ML,  an over-the-air analog-domain-based NN architecture is depicted in Fig.~\ref{model_fig:multiRIS}. This architecture comprises the following layers: one input layer, one convolutional (Conv) layer, real-to-complex (R2C) layer, one complex FC layer, one complex-to-real (C2R) layer, one real-valued FC layer, two complex activation layers (ReLU),  one complex batch normalization layer, one power normalization layer, one precoder layer, one RIS layer, one combiner layer, and one output layer.\footnote{Please kindly note that this work aims to demonstrate the feasibility of implementing a digital FC layer using a RIS-aided over-the-air transmission layer, thereby enabling a shift from digital-domain-based NN architectures to analog-domain-based counterparts. It is important to highlight that the proposed paradigm is generally applicable to any NN architecture, as long as it incorporates FC layers.} The Conv layer has $2$ output channels with a kernel size of  $3$, stride of $4$, and padding of $1$; 
the R2C layer converts the real-valued input into a complex-valued output with half the original dimensionality. Conversely, the C2R layer converts a complex-valued input into a real-valued output with twice the dimensionality.

We model the RIS-related channel using a Rician fading model, given by
\begin{align}
\!\!\!{\bf{G}}{\rm{ = }}\sqrt {\frac{K}{{K + 1}}} {{\bf{G}}_{{\rm{LoS}}}} + \sqrt {\frac{1}{{K + 1}}} {{\bf{G}}_{{\rm{NLoS}}}}, {\bf{G}} \in \left\{ {{{{\bf{\bar H}}}_i},{{{\bf{\hat H}}}_i}} \right\},
\end{align}
where $K$ represents the Rician factor, ${{\bf{G}}_{{\rm{LoS}}}}$  denotes the deterministic LoS component, and ${{\bf{G}}_{{\rm{NLoS}}}}$  
is a random component whose entries are independent and identically distributed complex Gaussian variables with zero mean and unit variance.

We evaluate performance using two datasets, namely the MNIST  and Fashion MNIST datasets, each containing $10$ classification categories and image dimensions of $28\times28$.  The number of training epochs and the batch size are set to $200$ and $32$, respectively.  In addition, each training batch is associated with one independent channel realization. Unless otherwise specified, the system parameters are set to  $N=49$ and  ${\sigma ^2} = 1$. 

\subsection{MNIST dataset}
In this subsection, we use the MNIST dataset to
verify our proposed scheme.
\subsubsection{Multi-RIS over-the-air computation}
We begin by training a digital-domain-based NN on the MNIST dataset as shown in Fig.~\ref{model_fig1}, and obtain the well-trained weight matrix $\bf W$ for the middle layer. Subsequently, we compute the over-the-air parameters, i.e., ${\bf F}_1$, ${\bf F}_2$, and $\bf \Theta$,  based on Algorithm~\ref{alg1}.

\begin{figure}[!t]
	\centerline{\includegraphics[width=3.3in]{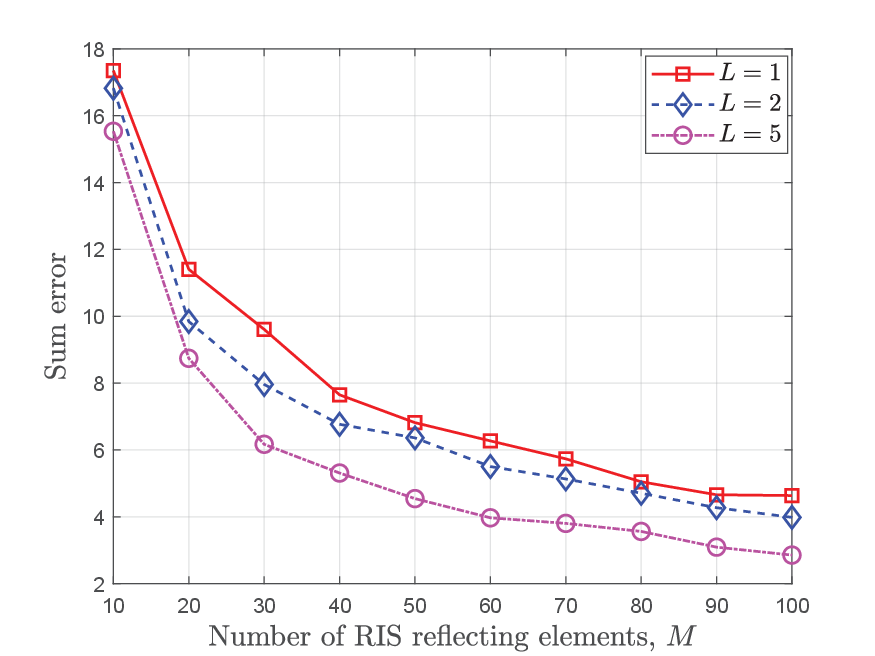}}
	\caption{Number of reflecting elements $M$ versus  sum imitation error for $L$ under $K=10~{\rm dB}$ and $P_{\rm max} = 10~{\rm dB}$.}  \label{SingleRISOAcal:fig11}
\end{figure}

Fig.~\ref{SingleRISOAcal:fig11} illustrates the relationship between the sum imitation error and the number of reflecting elements  $M$
 for various numbers of RISs $L=1, 2$, and $5$. 
 It can be observed that increasing 
 $M$ results in a lower imitation error due to the higher passive gain provided by the RIS. Additionally, increasing the number of RISs 
 $L$ also leads to a reduction in imitation error, particularly when 
 $M$ is large.
 For example, when  $M=100$, the sum imitation error is $2.8$ for $L=5$, compared to approximately 4.6 for $L=1$, which implies that about a $39.1\%$ reduction in imitation error.

\begin{figure}[!t]
	\centerline{\includegraphics[width=3.3in]{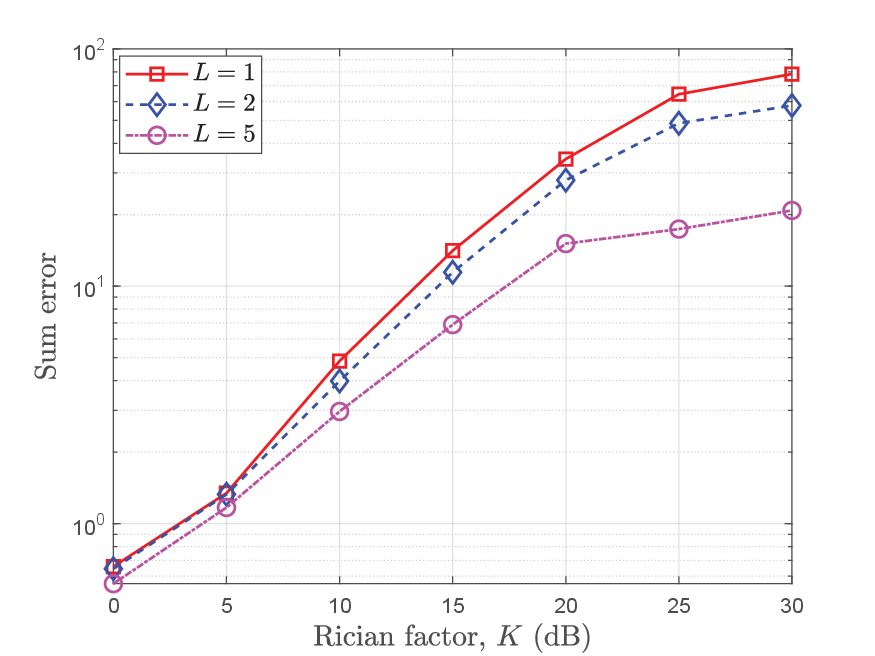}}
	\caption{Rician factor $K$     versus  sum imitation error for different $L$ under $M=100$ and $P_{\rm max} = 10~{\rm dB}$.}  \label{SingleRISOAcal:fig12}
\end{figure}
Fig.~\ref{SingleRISOAcal:fig12} investigates the impact of the Rician factor $K$ on the sum imitation error for different values of $L$. When $K$ is large, the channel becomes increasingly dominated by the LoS component, and the non-line-of-sight (NLoS) component becomes negligible.
This results in a rank-deficient channel,  significantly increasing the imitation error. However, deploying multiple RISs effectively creates additional LoS paths, thereby enhancing the channel rank and reducing the imitation error. 
 For example, when  $K = 30~{\rm dB}$, the sum imitation error for $L=5$ is $20.8$, compared to approximately  $78.3$ for $L=1$,  resulting in a  $73.4\%$ reduction in imitation error.
\begin{figure}[!t]
	\centerline{\includegraphics[width=3.3in]{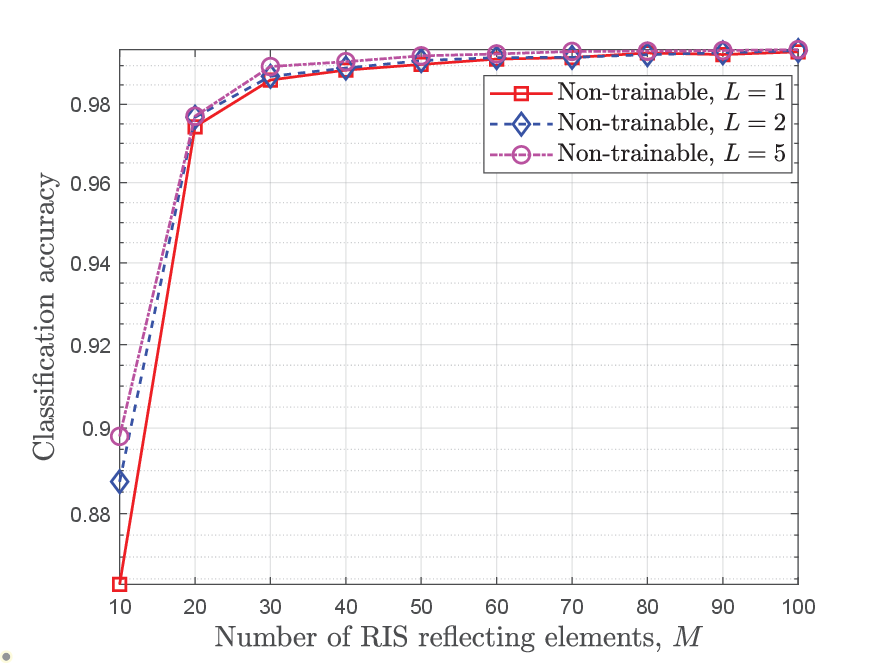}}
	\caption{$M$   versus  classification accuracy error for different $L$ under $K=10~{\rm dB}$ and $P_{\rm max} = 10~{\rm dB}$.}  \label{SingleRISOAcal:fig14}
\end{figure}

\begin{figure}[!t]
	\centerline{\includegraphics[width=3.3in]{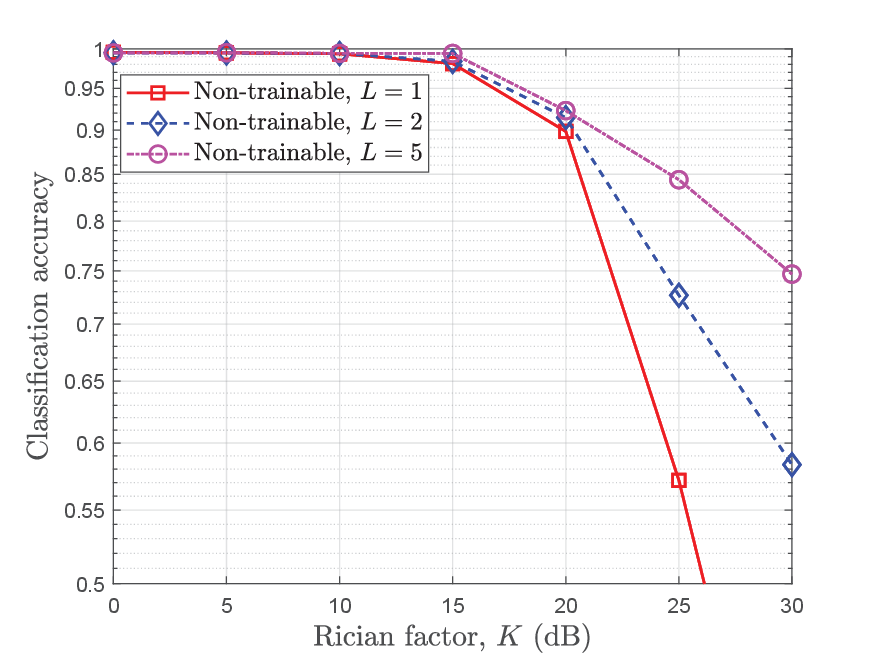}}
	\caption{  Classification accuracy as a function of the Rician factor $K$    for different $L$ under $M=100$ and $P_{\rm max} = 10~{\rm dB}$.}  \label{SingleRISOAcal:fig15}
\end{figure}
\subsubsection{Impact of over-the-air parameters on classification accuracy}
To  evaluate  the impact of  over-the-air parameters on  classification performance,  we first compute the parameters ${\bf F}_1$, ${\bf F}_2$, and $\bf \Theta$ using Algorithm~\ref{alg1}. The digital FC layer (i.e., the middle layer in Fig.~\ref{model_fig1}) is then replaced with these over-the-air components, and inference is performed over-the-air on test images from the MNIST dataset.

The influence of the parameters $M$ and $K$ on classification accuracy for various values of  $L$ is studied in  Fig.~\ref{SingleRISOAcal:fig14} and Fig.~\ref{SingleRISOAcal:fig15}, respectively.   Both figures demonstrate that a larger $L$ leads to higher classification accuracy. This result aligns with earlier findings in  Fig.~\ref{SingleRISOAcal:fig11} and Fig.~\ref{SingleRISOAcal:fig12}, where increased 
$L$ reduces the imitation error, which in turn improves classification performance.

\begin{figure}[!t]
	\centerline{\includegraphics[width=3.3in]{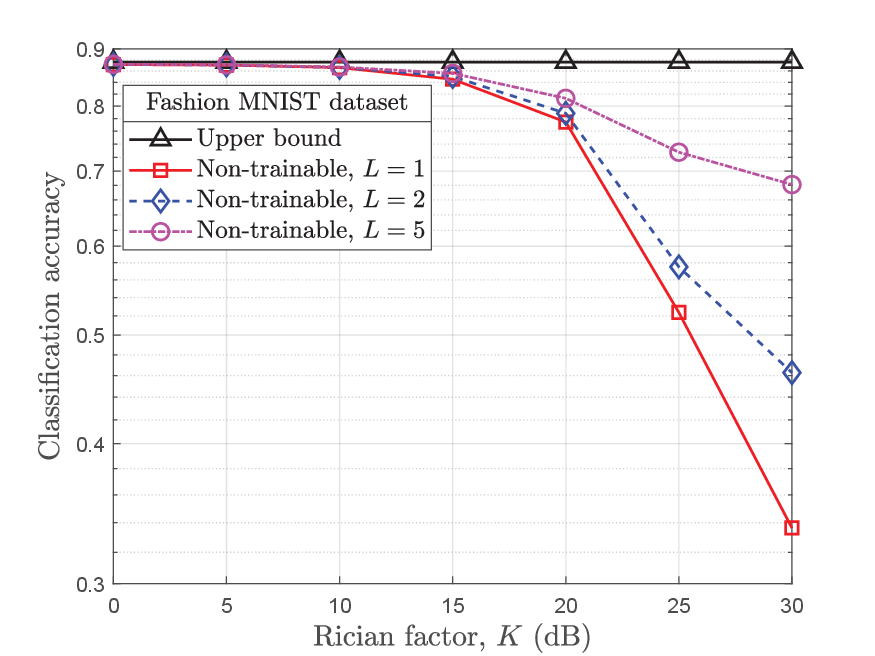}}
	\caption{Classification accuracy as a function of the Rician factor $K$ for the Fashion-MNIST dataset. }  \label{Fashion:fig18}
\end{figure}
\subsection{Fashion-MNIST dataset}
In Fig.~\ref{Fashion:fig18}, we examine the classification accuracy versus the Rician factor  $K$  for different values of  $L$, under $M=100$ and $P_{\rm max} = 10~{\rm dB}$, using the Fashion-MNIST dataset. The ``Upper bound" scheme is constructed using the original digital-domain-based architecture. As shown in  Fig.~\ref{Fashion:fig18}, 
the trend of the curves closely resembles that in Fig.~\ref{SingleRISOAcal:fig15}, indicating consistent behavior. This similarity is attributed to the spatial diversity gain enabled by the deployment of multiple RISs.

\section{Conclusion}
In this paper, we investigated RIS-aided MIMO systems for engineering the ambient wireless propagation environment to replicate digital FC layers through analog AirComp. To unveil the fundamental principles of RIS-enabled analog AirComp, we formulated and analyzed the imitation minimization error problem between a target digital FC layer and its analog counterpart. This was achieved by jointly optimizing the over-the-air parameters.
To tackle the resulting non-convex optimization problem, we proposed a low-complexity alternating optimization algorithm, where semi-closed-form and closed-form solutions were derived for all variable blocks. Simulation results demonstrated that the proposed AirFC scheme can be effectively implemented in the analog domain while achieving satisfactory classification accuracy.
Furthermore, our results revealed that employing a larger number of RISs can significantly enhance classification performance, especially in scenarios where the wireless channel is dominated by LoS components.

\bibliographystyle{IEEEtran}
\bibliography{AirFCRIS_conf}
\end{document}